\documentclass[12pt]{article}
\usepackage{curves}
\newcommand{\dd}{\mbox{\rm d}}
\newcommand{\DD}{\mbox{\rm D}}
\newcommand{\nnn}{\noindent}
\newcommand{\oo}{\over}
\newcommand{\p}{\partial}

\newcommand{\be}{\begin{equation}}
\newcommand{\ee}{\end{equation}}
\newcommand{\lbl}{\label}
\newcommand{\bi}{\bibitem}

\newcommand{\vs}{\vspace}

\newcommand{\qb}{\qbezier}

\def\bear{\begin{eqnarray}}
\def\ear{\end{eqnarray}}

\begin{document}
\baselineskip .7cm

\thispagestyle{empty}

\ 

\vs{1.5cm}

\begin{center}

{\bf \LARGE A Brane World Model with Intersecting Branes}

\vs{6mm}

Matej Pav\v si\v c\footnote{Email: MATEJ.PAVSIC@IJS.SI}

Jo\v zef Stefan Institute, Jamova 39, SI-1000 Ljubljana, Slovenia

\vs{1.5cm}

ABSTRACT
\end{center}

A brane world model is investigated, in which there are many branes
that may intersect and self intersect. One of the branes, being a
3-brane, represents our spacetime, while the other branes, if they intersect
our brane world, manifest themselves as matter in our 3-brane.
It is shown that such a matter encompasses dust of point particles and
higher dimensional $p$-branes, and all those objects follow "geodesics"
in the world volume swept by our 3-brane. We also point point out that
such a model can be formulated in a background independent way, and that the
kinetic term for gravity arises from quantum fluctuation of the brane.

\vs{1cm}

\nnn PACS: 1117, 0460 

\nnn KEY WORDS: Brane world, intersecting branes, background
independence, strings, $p$-branes

\vs{9mm}

\nnn Phone: +386 1 177-3780 ; Fax: +386 1 219-385  or 273-677

\newpage

\section{Introduction}

The idea that our world is a 3-brane embedded in a higher dimensional
bulk space is not new \cite{1}-\cite{2}. Recently it attracted much attention,
since Randall and Sundrum \cite{3} have found that gravity can be
localized on a brane. Such a property results as a solution to Einstein's
equations around a static 3-brane embedded in 5-dimensional space
with negative cosmological constant. Matter including the fields of
the standard model is confined to the brane, while gravity propagates
in the bulk, but in the Randall-Sundrum scenario gravity turns out
to be effectively localized on the brane too.

Besides the research exploring the properties of various "brane world"
scenarios based on the classical Einstein equations, there is also a lot
of activity aiming at formulating a consistent theory of quantum gravity.
In the works by Rovelli, Smolin and Baez \cite{4}, the need was
stressed that a really fundamental theory should be background independent.
This means that spacetime together with its metric should emerge from the
properties of some more basic objects. The basic objects could be spin
networks \cite{4a}, spin foams, or perhaps strings and various branes. A background
independent theory of $p$-branes should be formulated without using the
concept of a preexisting embedding space and metric. A configuration of
branes is all what exists in such an approach. There is no embedding space.
If there are many such branes, then they are supposed to form, up to a
good approximation, the embedding space. The latter space is in fact
identified with such a configuration of many branes.

We shall gradually build up the model. First we shall assume that we have 
a brane, representing a world (a brane world for short), moving in 
a background space $V_N$. Then we shall assume that $V_N$
is conformally flat. We shall take a special conformally flat metric,
such that it is singular on a set of branes. Then we shall observe that
the intersections of all those branes with some chosen brane behave as {\it
matter} on that brane. Such a matter consists of $p$-branes of various
dimensionalities and their equations of motions turn out to be those
of minimal surface (geodesic in particular, when $p=0$) in the brane
world metric. It is also possible that a brane world intersects itself.
Quantum fluctuations of such a brane induce the Einstein-Hilbert action
in the brane world and also quantum fluctuations of matter.

\vs{8mm}

\section{The brane in a bulk with a singular conformally flat metric}

\vs{2mm}

Let us consider a brane moving in a curved background embedding space
$V_N$, called {\it bulk}. Such a brane sweeps an $n$-dimensional surface
which I call {\it worldsheet}\footnote{
Usually, when $n>2$, such a surface is called {\it world volume}. Here I
prefer to retain the name {\it worldsheet}, by which we can vividly
imagine a surface in an embedding space.}.
The dynamical principle governing motion of the brane requires that its
worldsheet be a minimal surface. Hence the action is
\be
    I[\eta^a] = \int \sqrt{|{\tilde f}|} \, {\dd}^n x
\lbl{1}
\ee
where
\be
    {\tilde f} \equiv {\rm det} {\tilde f}_{\mu \nu} \; \; , \qquad 
    {\tilde f}_{\mu \nu} \equiv \p_{\mu} \eta^a \p_{\nu} \eta^b \gamma_{ab}
\lbl{2}
\ee
Here $x^{\mu}$, $\mu = 0,1,2,...,n-1$ are coordinates on the worldsheet
$V_n$, while $\eta^a (x)$ are the embedding functions. The metric of the
embedding space (from now on called also {\it bulk}) is $\gamma_{ab}$, and
the induced metric on $V_N$ is ${\tilde f}_{\mu \nu}$.

Suppose now that the metric of $V_N$ is conformally flat \cite{2,5}
(with $\eta_{ab}$ being the Minkowski metric tensor):
\be
    \gamma_{ab} = \phi \, \eta_{ab}
\lbl{3}
\ee
Then from (\ref{2}) we have
\be
    {\tilde f}_{\mu \nu} = \phi \, \p_{\mu} \eta^a \p_{\nu} \eta^b \eta_{ab}
    \equiv \phi f_{\mu \nu}
\lbl{4}
\ee
\be
     {\tilde f} \equiv {\rm det} {\tilde f}_{\mu \nu} = \phi^n \, {\rm det}
     f_{\mu \nu} \equiv \phi^n f
\lbl{5}
\ee
\be
     \sqrt{|{\tilde f}|} = \omega |f| \; , \quad \omega \equiv \phi^{n/2}
\lbl{6}
\ee
Hence the action (\ref{1}) reads
\be
      I[\eta^a] = \int {\dd}^n x \, \omega (\eta) \, \sqrt{|f|} {\dd}^n x
\lbl{7}
\ee
which looks like an action for a brane in a flat embedding space, except for
a function $\omega (\eta)$ which depends on position\footnote{
We use here the same symbol $\eta^a$ either for position coordinates in
$V_N$ or for the embedding functions $\eta^a (x)$.}
$\eta^a$ in the embedding space $V_N$.

Function $\omega (\eta)$ is related to the fixed background metric 
$\gamma_{ab}$ which is
arbitrary in principle. Let us now assume that $\omega (\eta)$ consists of
a constant part $\omega_0$ plus a singular part with support on a set
of surfaces $V_{m_j}$, of dimension $m_j$ and described by embedding
functions $\eta_j^a (x_j^{\mu_j})$, denoted $\eta_j$ for short:
\be
     \omega (\eta) = \omega_0 + \sum_j \int \kappa_j \delta^N (\eta - \eta_j)
      \, \sqrt{|f_j|} \dd x_j 
\lbl{8}
\ee
Here $\dd x_j \sqrt{|f_j|} \equiv {\dd}^{m_j} x_j \sqrt{|f(x_j)|}$ is the
invariant volume element on $V_{m_j}$.

Inserting (\ref{8}) into (\ref{7}) we obtain an action which contains
the kinetic term for the worldsheet $V_n$ and an interactive term between
$V_n$ and $V_{m_j}$:
\be
     I[\eta] = \int \omega_0 \sqrt{|f|} \, \dd x + 
     \sum_{j} \int \kappa_j \, \delta^N (\eta - \eta_j)
     \sqrt{|f|} \sqrt{|f_j|}\, \dd x \, \dd x_j
\lbl{8a}
\ee
The set of worldsheets $V_{m_j}$ form a background with the singular 
conformally flat metric, given by (\ref{3}),(\ref{6}) and (\ref{8}), in
which the worldsheet $V_n$ lives.

\section{A system of many intersecting branes}

Now we shall assume that $V_{m_j}$ are dynamical too. Therefore we add
a corresponding kinetic term to the action. So we obtain an action for
a system of intersecting branes $\eta_i$, $i=1,2,...$:
\be
    I[\eta_i] = \sum_i \int \omega_0 \sqrt{|f_i|} \, \dd x_i + {1\oo 2} 
    \sum_{i j} \int \omega_{i j} \, \delta^N (\eta_i - \eta_j)
    \sqrt{|f_i|} \sqrt{|f_j|}\, \dd x_i \dd x_j
\lbl{9}
\ee
Besides the kinetic term for free branes, our action (\ref{9}) contains
also the interactive terms. The interactions result from the intersections of
the branes.

The equations of motion for the $i$-th brane are
\be
    \p_{\mu} \left [\sqrt{|f_i|} \p^{\mu} \eta_i^a \left ( \omega_0 +  
    \sum_{i\neq j} \int \omega_{i j} \, \delta^N (\eta_i - \eta_j)
     \,  \sqrt{|f_j|} \dd x_j \right ) \right ] = 0 
\lbl{10}
\ee
The same equations (with the identification $\eta \equiv \eta_i$, $\kappa_j
\equiv \omega_{ij}$) follow also from (\ref{8a}). However, with (\ref{9}) we
have a self consistent system, where each brane determines the motion of all
the others.

Returning now to the action (\ref{8a}) experienced by one of the branes
whose worldsheet is represented by $\eta_i^a (x_i) \equiv \eta^a (x)$,
we find after integrating out $x_j$, $j \neq i$ that 
\be
    I[\eta] = \omega_0 \int \dd^n x \, \sqrt{|f|} + \sum_j \kappa_j \int
    \dd^n x \, {\dd}^{p_j + 1} \xi \, ({\rm det} \, \p_A X_j^{\mu} 
    \p_B X_j^{\nu}
    f_{\mu \nu} )^{1/2} \, \delta^n (x - X_j (\xi))
\lbl{11}
\ee
For various $p_j$, the latter expression is an action for a system of point
particles $(p_j = 0)$, strings $(p_j = 1)$, and higher dimensional branes
$(p_j = 2,3,...)$, described by $X_j^{\mu} (\xi)$,
moving in the background metric $f_{\mu \nu}$, which is
the induced metric on our brane $V_n$ (see Fig.1)

\setlength{\unitlength}{.5mm}
\begin{picture}(120,130)(-60,0)

\thicklines
\qbezier(10,60)(38,67)(60,81)
\qbezier(60,81)(64,55)(61,30)
\qbezier(19,12)(15,50)(10,60)
\qbezier(19,12)(29,20)(61,30)
\qbezier(32,12)(47,26)(81,26)
\qbezier(81,26)(84,33)(85,50)
\qbezier(85,50)(85,72)(91,86)
\qbezier(91,86)(63,81)(32,93)
\qbezier(32,93)(29,79)(30,66.4)
\qbezier(32,12)(33,17)(33,19.5)
\qbezier(84.4,40)(95,42.5)(103,48)
\qbezier(103,48)(96,83)(101,115)
\qbezier(101,115)(86,98)(66,85)

\put(64,54){$V_{p+1}$}
\put(40,95){$V_n$}
\put(103,111){${\hat V}_m$}
\put(125,76){$V_N$}

%\put(1,5){\framebox(6,2){\large $a^2 + b^2 = c^2$}}
%\put(3.5,5){\vector(0,-1){5}}
%\put(7,3){$\int d x e^{i p x}$}
%\put(6,3){$\Biggl[$}
%\put(5.5,3){\vector(-1,0){2} }
%\put(0,0){\framebox(15,14)}

\end{picture}

\begin{figure}[h]
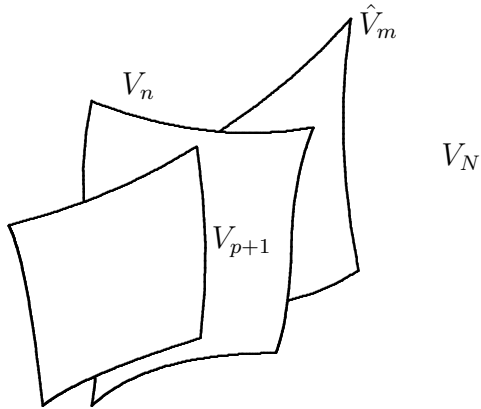

\caption{The intersection between two different branes $V_n$ and ${\hat V}_m$
can be a $p$-brane $V_{p+1}$.}
\end{figure}

We see that the interactive term in (\ref{9}) manifests itself in
various ways, depending on how we look at it. It is a manifestation of the
fact that the metric of the embedding space is curved (in particular, 
the metric is singular on the system of branes). From the point of view of
a chosen brane $V_n$ the interactive term becomes the action for a system of
$p$-branes (including point particles) moving on $V_n$. If we now adopt
the {\it brane world} view, where $V_n$ is our spacetime, we see {\it that
matter on $V_n$ comes from other branes's worldsheets which happen to intersect 
our worldsheet $V_n$.} Those other branes, in turn,  are responsible 
for the non trivial metric of the embedding space.

\subsection{The brane interacting with itself}

In (\ref{8a}) or (\ref{9}) we have a description of a brane
interacting with other branes. What about self interaction? In the second term
of the action (\ref{8a}),(\ref{9}) we have excluded self interaction.
In principle we should not exclude self interaction, since there is no reason why
a brane could not interact with itself.

Let us return to the action (\ref{8a}) and let us calculate 
$\omega (\eta)$,\
this time assuming for simplicity that there is only one brane $V_{m_j}
\equiv {\hat V}_m$ which coincides with our brane $V_n$. Hence
the intersection is the brane $V_n$ itself, and according to (\ref{8})
we have
\bear    
    \omega (\eta) &=& \omega_0 + \kappa \int {\dd}^n {\hat x} \sqrt{|{\hat f}|}
    \, \delta^N (\eta - {\hat \eta} ({\hat x})
    \nonumber \\
     &=& \omega_0 + \kappa \int {\dd}^n \xi \, \sqrt{|{\hat f}|} \, 
    \delta^n (x - X(\xi)) \nonumber \\ 
   &=& \, \omega_0 + \kappa \int {\dd}^n x \, \delta^n (x - X(x)) =
      \omega_0 + \kappa
\lbl{III1.39b}
\ear  
Here the coordinates $\xi^A$, $A = 0,1,2,..., n-1$ cover the manifold $V_n$,
and ${\hat f}_{AB}$ is the metric of $V_n$ in coordinates $\xi^A$. The other
coordinates are $x^{\mu}$, $\mu = 0,1,2,...,n-1$. In the last step in
(\ref{III1.39b}) we have used the property that the measure is invariant,
${\dd}^n \xi \sqrt{|{\hat f}|} = {\dd}^n x \sqrt{|f|}$.

The result (\ref{III1.39b}) demonstrates that we do not need to separate a
constant term $\omega_0$ from the function $\omega (\eta)$. For a brane moving
in a background of many branes we can replace (\ref{8}) with
\be
      \omega (\eta) = \sum_j \int \kappa_j \delta^N (\eta - \eta_j)
      \, \sqrt{|f_j|} \dd x_j 
\lbl{III1.39c}
\ee
where $j$ runs over {\it all} the branes within the system. Any brane feels
the same background, and its action for a fixed $i$ is
\be
   I[\eta_i] = \int \omega (\eta_i) \sqrt{|f_i|} \dd x_i = 
   \sum_j \int \kappa_j \delta^N (\eta_i - \eta_j)
   \, \sqrt{|f_i|} \sqrt{|f_j|} \dd x_i \, \dd x_j
\lbl{III1.39d}
\ee
However the background is self consistent: it is a solution to the 
variational principle given by the action
\be
    I[\eta_i] = \sum_{i \ge j} \omega_{ij} \delta^N (\eta_i - \eta_j) 
    \sqrt{|f_i|} \sqrt{|f_j|} \dd x_i \, \dd x_j
\lbl{III1.39e}
\ee
where now also $i$ runs over {\it all} the branes within the system;
the case $i = j$ is also allowed.

In (\ref{III1.39e}) the self interaction or self coupling occurs whenever
$i = j$. The self coupling term of the action is ($\kappa_i \equiv 
\omega_{ii}$)
\begin{eqnarray}
      I_{\rm self} [\eta_i] &=& \sum_i \kappa_i \int \delta^N (\eta (x_i) -
      \eta_i (x'_i)) \sqrt{|f_i (x_i)|} \sqrt{|f_i (x'_i)|} \dd x_i \dd x'_i
      \nonumber \\
      &=& \sum_i \kappa_i \int \delta^N (\eta - \eta_i (x_i)) 
      \delta^N (\eta - \eta_i (x'_i))
      \sqrt{|f_i (x_i)|} \sqrt{|f_i (x'_i)|} \dd x_i \dd x'_i {\dd}^N \eta
      \nonumber \\
      &=& \sum_i \kappa_i \int \delta^N (\eta - \eta_i (x_i)) \delta^{n_i}
      (x_i - x'_i) \sqrt{|f_i (x_i)|} \dd x_i \dd x'_i {\dd}^N \eta
      \nonumber \\
      &=& = \sum_i \kappa_i \, \sqrt{|f_i (x_i)|} {\dd}^{n_i} x_i
\lbl{III1.39f}
\ear
where we have used the same procedure which led us to eq.(\ref{III1.39b}). 
We see that the interactive action (\ref{III1.39e})
automatically contains also the minimal surface terms, so that they do
not need to be postulated separately.

\subsection{A system of many branes creates bulk and its metric}

We can now imagine that a system of
branes (a brane configuration) can be identified with the embedding space
in which a single brane moves. Here we have a concrete realization of that 
idea. We have a system of branes which intersect. The only interaction between
the branes is due to intersection ( "contact interaction"). The interaction at
the intersection influences the motion of a (test) brane: it feels a potential
because of the presence of other branes. If there are many branes and a test
brane moves in the midst of them, then on average it feels a metric field
which is approximately continuous. Our test brane moves in an effective metric
of the embedding space.

A single brane or several branes give the singular conformal metric. Many 
branes are expected to give, on average, an arbitrary metric.

\setlength{\unitlength}{.5mm}
\begin{picture}(120,130)(-30,0)

\thicklines
\qb(20,92)(50,81)(105,116)
\qb(105,116)(96,94)(100,51)
\qb(100,51)(84,39)(63,34)
\qb(63,34)(44,29)(26,15)
\qb(26,15)(14,60)(21,92)

\qb(27,113)(79,79)(131,93)
\qb(131,93)(123,87)(116,34)
\qb(116,34)(65,48)(39,42)
\qb(39,42)(47,65)(27,113)

\qb(80,72)(120,101)(125,120)
\qb(125,120)(121,95)(123,47)
\qb(123,47)(89,32)(76,19)
\qb(76,19)(80,31)(80,72)

\qb(78,123)(100,99)(139,96)
\qb(139,96)(134,72)(135,42)
\qb(135,42)(91,58)(67,55)
\qb(67,55)(76,74)(78,123)

%\put(1,5){\framebox(6,2){\large $a^2 + b^2 = c^2$}}
%\put(3.5,5){\vector(0,-1){5}}
%\put(7,3){$\int d x e^{i p x}$}
%\put(6,3){$\Biggl[$}
%\put(5.5,3){\vector(-1,0){2} }
%\put(0,0){\framebox(15,14)}

\end{picture}

\begin{figure}[h]
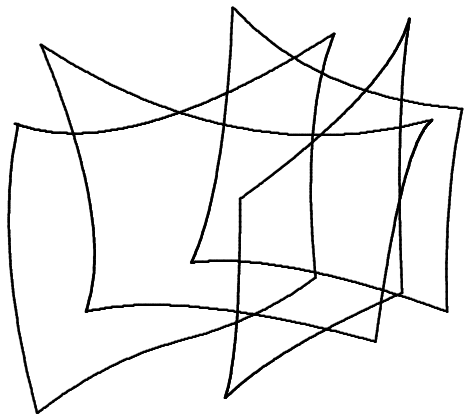

\caption{A system of many intersecting branes creates the bulk metric. In the
absence of the branes there is no bulk (no embedding space).}
\end{figure}

There is a close interrelationship between the presence of branes and
the bulk metric. In the model we discuss here the bulk metric is singular on
the branes, and zero elsewhere. Without the branes there is no metric and no
bulk. Actually the bulk consists of the branes which determine
its metric.

Something quite analogous occurs in string theory, more precisely, in
the theory of closed superstrings. Although classically the string theory
is formulated in a background spacetime with a fixed (Minkowski) metric,
it turns out after quantization that the background metric in which the
quantum string moves cannot be fixed. The string itself determines
what are the equations of motion for the metric of the embedding space
(the target space in the string theoretic jargon). This could be
intuitively understood by noticing that the quantum string automatically
involves many strings. A generic quantum state is a many strings state
and effectively it leads to gravity in target space. What is still not
quite satisfactory in string theory is its background dependent starting
point, namely use of the Minkowski metric. I believe that the many
intersecting brane model (which, of course, includes also strings)
resolves the issue of background independence at the classical level, since
the action (\ref{III1.39e}), which includes also self interaction, contains no
metric of a background embedding space. It is true that in (\ref{III1.39e})
we have the quantity $\sqrt{f_i}$, $f_i \equiv {\rm det} \, {f_i}_{\mu \nu}$,
where ${f_i}_{\mu \nu} \equiv \p_{\mu} \eta_i^a \p_{\nu} \eta_i^b \eta_{ab}$.
The latter quantity is the metric on the $i$-th brane worldsheet, but now
it can not be considered as a metric induced from an embedding space metric,
since according to (\ref{III1.39c}), (\ref{3}) and (\ref{6})
 the latter metric vanishes outside the branes. Hence in
effect there is no embedding space, apart from the system of branes itself.
The fixed quantity $\eta_{ab}$ can even  less be interpreted as a metric 
of an embedding space. It is the Minkowski metric of the flat space to which
there corresponds a conformally flat space which, because of the singular
conformal factor, is identified with the system of branes. The latter
conformally flat space is our dynamical system, but the former
flat space (with metric $\eta_{ab}$) is not, and therefore can hardly be
considered as a background space for our dynamical system of branes.

\section{The origin of matter in brane world}

Our principal idea is that we have a system of branes (a brane configuration).
With all the branes in the system we associate
the embedding space (bulk). One of the branes (more precisely, its worldsheet)
represents our spacetime. Interactions between the branes (occurring at the
intersections) represent {\it matter} in spacetime.\

\subsection{Matter from the intersection of our brane with other branes}

We have seen that matter in $V_n$ naturally occurs as a result of the
intersection of our worldsheet $V_n$ with other worldsheets. We obtain
exactly the stress-energy tensor for dust of point-particles, or $p$-branes
in general. Namely, varying the action (\ref{11}) with respect to
$\eta^a (x)$ we obtain
\be
    \omega_0 {\DD}_{\mu} {\DD}^{\mu} \eta_a + {\DD}_{\mu} (T^{\mu \nu} \p_{\nu}
    \eta_a) = 0
\lbl{III1.40}
\ee
with
\be
    T^{\mu \nu} = \sum_j \int \int {\dd}^{p_j +1} \xi \, \, ({\rm det} \, \p_A
    X_j^{\mu} \p_B X_j^{\nu} \, f_{\mu \nu})^{1/2} \, {{\delta^n (x - X_j(\xi))}
    \oo {\sqrt{|f|}}}
\lbl{III1.41}
\ee
being the stress-energy tensor for a system of $p$-branes (which are the
intersections of $V_n$ with the other worldsheets). By ${\DD}_{\mu}$
we denote the covariant derivative with respect to the world sheet
metric $f_{\mu \nu} \equiv \p_{\mu} \eta^a \p_{\nu} \eta_a$.
The above expression
for $T^{\mu \nu}$ holds if the extended objects have any dimensions $p_j$.
In particular, when {\it all} objects have $p_j = 0$ (point-particles)
eq.(\ref{III1.41}) becomes
\be
     T^{\mu \nu} = \sum_j \kappa_j \int \dd \tau \, {{{\dot X}^{\mu} {\dot X}^{
     \nu}}\oo {\sqrt{{\dot X}^2}}} \, {{\delta (x - X(\tau))}\oo {\sqrt{|f|}}}
\lbl{III1.42}
\ee

From the equations of motion (\ref{III1.40}) we obtain\footnote{
We contract (\ref{III1.40}) by $\p_{\nu}\eta^a$ and take into account the
identity ${\DD} _{\alpha} {\DD}_{\beta} \eta_a \, \p_{\nu}\eta^a = 0$.}
\be
    {\DD}_{\mu} T^{\mu \nu} = 0
\lbl{III1.42a}
\ee
which implies that any of the objects sweeps a minimal surface
$V_{p+1}$ in $V_n$. When $p_j = 0$ we have a geodesic in $V_n$.
                 
\subsection{Matter from the intersection of our brane with itself}

Our model of intersecting branes allows for the possibility that a brane
intersects itself, as schematically illustrated in Fig. 3.

\setlength{\unitlength}{1mm}
\begin{picture}(120,110)(-15,0)
\thicklines

\qb(14,69)(33,40)(51,46)
\qb(51,46)(58,48)(57,55)
\qb(57,55)(56.5,62)(44,60)
\qb(44,60)(24,53)(47,40)
\qb(47,40)(55,36)(66,34)
\qb(66,34)(65,17)(60,0)
\qb(14,69)(14,43)(8,18)
\qb(8,18)(33,16)(60,0)
\qb(38.2,46.2)(40,30)(33,13)
\qb(57,52)(55,45)(54,37)
\curve(34.7,52, 34.6,48)
\put(27,31){$V_{p+1}$}
\put(52,25){$V_n$}
\put(70,50){$V_N$}

\end{picture}

\begin{figure}[h]
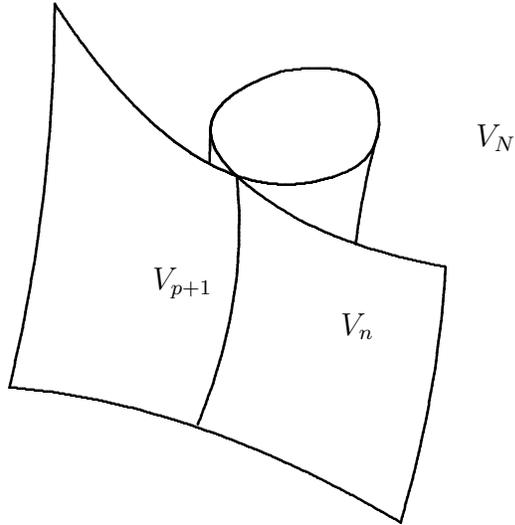

\caption{
Illustration of a self-intersecting brane. At the intersection $V_{p+1}$,
because of the contact interaction, the stress-energy tensor on the brane $V_n$
is singular, and it manifests itself as matter on $V_n$. The manifold
$V_{p+1}$ is a worldsheet swept by a $p$-brane and it is a minimal surface
(e.g. a geodesic, when $p=0$) in $V_n$.}

\end{figure}

The analysis used so far is valid also for the situations like the one
in Fig. 3, if we divide the worldsheet $V_n$ in two pieces which are glued
together at a submanifold $C$, situated somewhere within the "loop" region.

There is a variety of ways a worldsheet can self intersect. For instance,
It may intersect itself many times to form a sort of helix or spiral. Instead 
of the intersection with a single loop, like in Fig. 4, the intersection 
may form a double or triple loop (Fig. 4).

\setlength{\unitlength}{1mm}
\begin{picture}(120,140)(10,-20)
\thicklines

\qb(23,111)(43,73)(72,72)
\qb(72,72)(92,70)(91,86)
\qb(91,86)(91,96)(78,93)
\qb(78,93)(67,90)(69,80)
\qb(69,80)(71,68)(85,67)
\qb(85,67)(106,66)(99,86)
\qb(99,86)(93,100)(80,100)
\qb(80,100)(62,100)(60,88)
\qb(60,88)(56,71)(82,60)
\qb(82,60)(94,54)(106,69)
\qb(106,69)(111,76)(104,92)
\qb(104,92)(98,105)(80,106)
\qb(80,106)(57,108)(52,91)
\qb(52,91)(46,71)(85,47)
\qb(85,47)(95,40)(111,36)
\qb(111,36)(108,16)(110,-10)
\qb(110,-10)(70,-4)(51,16)
\qb(23,111)(27,80)(17,55)
\qb(17,55)(28,28)(51,16)

\qb(107.5,72)(102,53)(103,39)
\qb(90.8,81)(89.5,76)(90,67.5)
\qb(100,73)(98,69)(97.8,61.7)
\qb(68.7,82)(69,77)(68.6,72.6)
\qb(59.4,83)(60,80)(59.5,74.6)
\qb(51.3,87)(51,82)(50.4,78.8)

\linethickness{.5mm}
\qb(72.6,72)(72.9,68)(72.5,65.2)
\qb(62.8,73)(63.2,69)(62,64.4)
\qb(52.8,77.7)(56,45)(51,16)
\put(72.6,72){\circle*{2}}
\put(62.8,73){\circle*{2}}
\put(52.8,77){\circle*{2}}

\end{picture}

\begin{figure}[h]
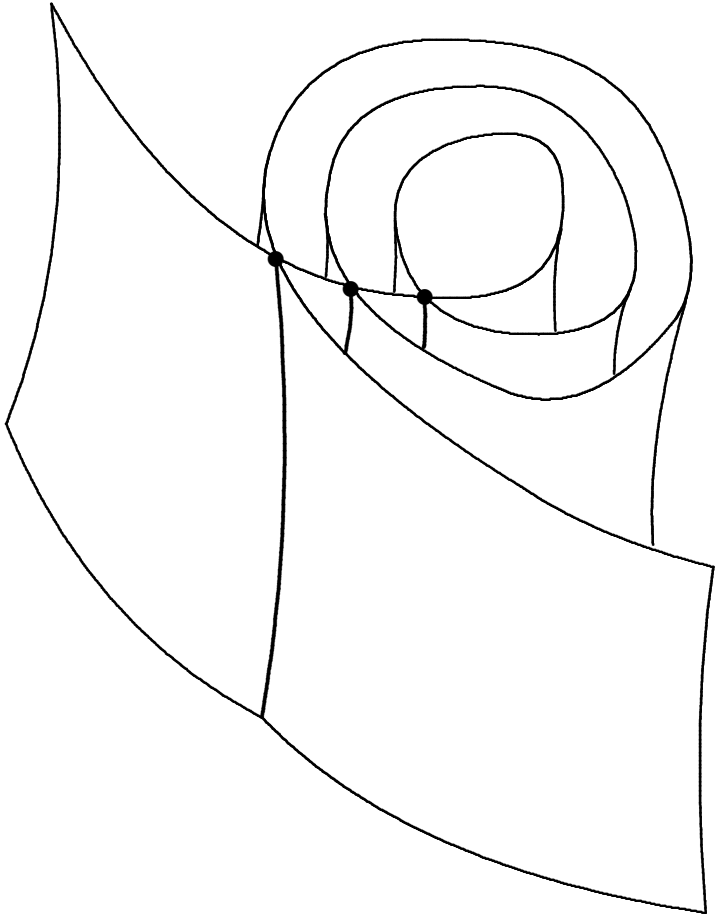

\caption{Example of a worldsheet intersecting with itself in a triple loop.
}

\end{figure}

In this respect some interesting new possibilities occur, waiting to be
explored in detail. For instance, it is difficult to imagine how the
three particles entangled in the topology of the situation in Fig. 4
could be separated to become asymptotically free. Hence this might be
a possible classical model for hadrons composed of quarks; the extra dimensions
of $V_n$ would bring, via Kaluza-Klein mechanism, the chromodynamic force
into the action. Moreover, the topology of the model appears to be chiral.

In summary, it is obvious that a self intersecting brane can provide a variety
of matter configurations on the brane. This is a fascinating and
intuitively clear mechanism for the origin of matter in a brane world.

\section{Discussion and conclusion}

The curvature scalar does not occur in the brane world action
(\ref{9}). In previous publications \cite{5,7} we have noticed that
the Einstein-Hilbert action on the brane's worldsheet can be induced 
from quantum fluctuations of the brane. A similar model had also been
considered within the idea of Sakharov's {\it induced gravity} \cite{6}
in refs.\cite{6a}. Instead of the action (\ref{9}) it is convenient to take
another, classically equivalent action, which is not only a functional
of the embedding functions $\eta^a (x)$, but also of the induced metric
$g_{\mu \nu}$. Such an action is known under the name {\it sigma model
action} or the Howe-Tucker action \cite{6b}. 
The quantization of the latter action
enables one to express {\it an effective action} as a functional
of $g_{\mu \nu}$. The effective action is obtained in the Feynman
functional integral in which we functionally integrate over $\eta^a (x)$,
so that what remains is a functional dependence on the induced metric
$g_{\mu \nu}$. This effective action contains the Ricci scalar $R$ and
its higher orders. Therefore the field equations contain the Einstein
equations on the brane worldsheet and the terms arriving from higher orders
of $R$. In other words, in the effective theory obtained after performing
the quantum average over various branes with the same induced metric
$g_{\mu \nu}$, the latter approximately satisfies the Einstein equations.
If having not a single brane action but an action, like (\ref{9}), for a 
system of many branes which can intersect, and self intersect,
then we obtain on a chosen brane the {\it matter} term for point-particle 
and higher $p$-brane sources. Quantum fluctuations of the 3-brane 
render the state of those matter sources to behave as
quantum sources. In the case of (bosonic) point particles the latter source
turns out to be  just the usual action for a scalar field \cite{5,7}. A
generalization to fermionic branes and hence to fermionic sources on
the world brane has not yet been explicitly constructed, but I expect it
should be straightforward, starting from the existing knowledge of superstrings
and supersymmetry.

Nowadays there is a lot of activity in the so called "brane world" scenario,
but its full power seems not yet been entirely appreciated. Conventionally,
strings and higher $p$-branes are considered as extended relativistic
objects in spacetime which necessarily has more than 4 dimensions (e.g. 26
for bosonic strings). Then there arises a problem of how to compactify all
those extra dimensions. Various ingenious models and methods are being
investigated. I prefer to adopt an alternative view, namely, that a
4-dimensional worldsheet swept by a 3-brane already represents spacetime.
Hence,
no compactification of dimensions of the embedding space (called also the
target space or bulk) $V_N$ is needed, since the latter space is not
our spacetime. Moreover, in effect an embedding space was shown to be identical
with the system of many branes. One of those branes is our world
(brane world), while
the other branes, if they intersect our brane, are manifested as {\it matter} 
in our world. So the other branes can have a physical influence on our
world as well.

\vs{8mm}

\centerline{\bf Acknowledgement}

\vs{5mm}

This work was supported by the Slovenian Ministry
of Science and Technology under Contract J1-7455.

\end{document}